\documentclass{aa}
\usepackage{graphicx}
\begin{document}
  \title{Anisotropy in the angular distribution of the long gamma-ray bursts?}

  \author{A. M\'esz\'aros \and J. \v{S}to\v{c}ek}

   \offprints{A. M\'esz\'aros}

   \institute{Astronomical Institute of the Charles University,
              V Hole\v{s}ovi\v{c}k\'ach 2, 180 00 Prague 8,
          Czech Republic\\
              \email{meszaros@mbox.cesnet.cz, jirsto@seznam.cz}       }
   \date{Received March 15, 2002; accepted  ..........}

   \abstract{
   The gamma-ray bursts detected by the BATSE instrument may be separated
into ``short", ``intermediate" and ``long" subgroups. Previous
statistical tests found an anisotropic sky-distribution on large
angular scales for the intermediate subgroup, and probably also
for the short subgroup. In this article the description and the
results of a further statistical test - namely the nearest
neighbour analysis - are given. Surprisingly, this test gives an
anisotropy for the long subgroup on small angular scales. The
discussion of this result suggests that this anisotropy may be
real.
}

   \maketitle

\section{Introduction}

The separation of the gamma-ray bursts (hereafter GRBs) - detected
by the BATSE instrument (\cite{mee00}) - into subgroups is done
(\cite{kou}, \cite{ho98}, \cite{muk98}) with respect to $T_{90}$,
the duration during which  90\% of the radiation of a burst is
measured). Bursts are either short ($T_{90} \leq 2$ s), or
intermediate ($2$ s $ < T_{90} \leq 10$ s), or long ($T_{90}
> 10$ s). Nowadays it is practically sure that the long and short
subgroups are different phenomena (\cite{nor01}, \cite{hor01}).
The si\-tuation concerning the intermediate subgroup is unclear;
some authors query even the reality of this subgroup itself
(\cite{HA}). From the afterglow data the cosmological origin is
directly confirmed for the long bursts only. They are usually at
high redshifts. For the short and intermediate GRBs there is only
indirect evi\-dence for a cosmological origin, and concrete
redshifts are unknown (for a survey of these questions see
\cite{me01}).

During the last years one of the authors together with various
collaborators provided several statistical tests verifying the
isotropy in the angular distribution of GRBs. These tests were
based on the binomial distribution (\cite{me97}, \cite{bal98},
\cite{bal99}), on spherical harmonics (\cite{me00a}), on the
counts-in-cells method (\cite{me00b}), on the two-point angular
correlation function (\cite{me00c}), and on multifractal methods
(\cite{va01}). These tests (for a summary see \cite{mea01}) give
an anisotropy for the intermediate subgroup. The short subgroup
also seems to be distributed anisotropically; nevertheless, there
are only a few tests that reject isotropy at a high enough
confidence level. The long subgroup seems to be distributed
isotropically; here only the test based on the two-point angular
correlation function rejects the null hypothesis of isotropy
(\cite{me00c}). Recently and fully independently these results
were confirmed by Litvin et al. (2001).

In this paper we present the results of a new test; namely of the
nearest neighbour analysis (hereafter NNA). This test
(\cite{ST89}) is a standard statistical test, and - as far as
known - has not been used yet for GRBs.

The paper is organized as follows. In Sect. 2 the method is described.
Sect. 3 compares this method with other methods.
Sect. 4 presents the results of the test. Sect. 5 discusses and summarizes
the conclusions of the paper.

\section{The method}

NNA is a standard statistical test, which compares the actual
angular distances among the objects on the surface of a sphere
having unit radius with the theoretical angular distances in a
randomly and isotropically distributed sample.

The theory of NNA was formulated by Scott \& Tout (1989). Here
this theory is only recapitulated and specified for the sky
distribution of GRBs.

Let there be $N$ objects on the sky, which are distributed
randomly. $N$ should be $\geq 2$; for our purpose we may assume 
$N \gg 1$. We arbitrarily choose one object (``first object"). At an
angular distance $\beta$ ($0 \leq \beta \leq \pi$) from the first
object we define an infinitesimal belt with thickness $d\beta$.
This belt is defined by the distance interval $[\beta, (\beta +
d\beta)]$. The probability of having  $(L-1)$ objects at the
distance $\leq \beta$, and one object in the infinitesimal belt is
given by
$$
p_{L}(\beta) d \beta =\frac{(N-1)!}{2^{N-1}(N-L-1)!(L-1)!}\times
$$
\begin{equation}
\sin\beta (1-\cos\beta)^{L-1} (1+\cos\beta)^{N-L-1} d \beta .
\end{equation}
$L$ can be $=1, 2, ..., (N-1)$. For $L=1$ the nearest (or: the
first nearest) object lies in  the belt, for $L=2$ the second
nearest object lies in the belt, etc.

The integral probability $\int_{0}^{\beta} p_L(\beta') d\beta' =
P_L (\beta)$ defines the probability that there are $L$ and
exactly $L$ objects in the neighbourhood of the first object; the
neighbourhood is defined by the distances 
$\leq \beta$. $P_L(\pi) = 1$, as it should be. For $L=1$ and $L=2$ one
obtains
\begin{equation}
P_{1}(X) = 1-\left(1-\frac{X}{N-1}\right)^{N-1},
\end{equation}
and
$$
P_{2}(X_2) =
1-\left(1-\frac{X_2}{(N-1)(N-2)}\right)^{N-1} -
$$
\begin{equation}
\frac{X_2}{N-2}\left(1-\frac{X_2}{(N-1)(N-2)}\right)^{N-2},
\end{equation}
respectively, where $X = (N-1) \sin^2 (\beta/2)$, and
$X_2=(N-1)(N-2)\sin^2\frac{\beta}{2}$.
The introduction of $X$ and $X_2$ - instead of $\beta$ - simplifies
the formulas (\cite{ms97}).

If the integral probabilities for different $L$ are given by these
analytical formulas, we can use them as theoretical cumulative
probability distributions. The standard Kolmogorov-Smirnov test
can be used to compare them with the measured empirical cumulative
distributions (\cite{pre}; Chapt.14.3).

The test for $L=1$ should be done as follows. The theoretical
function $P_1(X)$ is defined for $0 \leq X \leq (N-1)$ and is
monotonously increasing from $0$ to $1$. The measured $(N-1)$
first nearest neighbour distances are sorted into increasing
sequence. Then, for any distance in this sequence, one obtains a
value $X$, and at this value the empirical cumulative distribution
function is increased by the value $(N-1)^{-1}$. Hence the
empirical cumulative function also runs from $0$ to $1$ for the
same range of $X$. The maximal absolute value $D$ between the two
different cumulative distribution functions defines the
significance level for given $(N-1)$ (\cite{pre}; Chapt.14.3).

One can do $(N-1)$ tests, because the test may be done for any
allowed $L$. For any $L$ the theoretical cumulative function can
be calculated  and is an analytical function. The $(N-1)$ $L$-th
nearest neighbour measured angular distances can also be obtained
from the positions of objects.

However, one does not need to provide all possible $(N-1)$ tests
for one sample. Let us assume that the test is made for $L=1$,
then for $L=2$, ..., then for $L = (N-1)$. If one assumes that the
null hypothesis is true, viz. that the objects are distributed
randomly on the surface of the sphere, then no test should reject
this hypothesis. Nevertheless, it is unlikely that going to bigger
and bigger $L$ will give any new result. This may be seen as
follows.

There are $N(N-1)/2$ distances among the observed $N$ objects. But
only $(2N-3)$ distances are independent. ($N$ objects on the
sphere are defined by $2N$ coordinates [any object is defined by
two spherical angles, e.g. by $\vartheta$ and $\varphi$].
Nevertheless, one object may be taken - without loss of generality
- at the pole [i.e. one may have $\vartheta = 0$ for this ``first
object"]. A second arbitrary object may still be at $\varphi = 0$.
There are $(N-1)$ independent distances known immediately: they
are the $\vartheta$ coordinates of $(N-1)$ objects giving the
distances from the first object. There are a further $(N-2)$
independent distances from the second object. They can be
calculated  if the $\varphi$ coordinates are known. Then any
further distances can be calculated from these $(2N-3)$
independent distances.)

This means that, if one uses only $L=1$ and $L=2$, then one uses
$2N$ distances in these two tests, which is practically identical
to the number of independent distances. (For $N \gg 1$ the
difference between $2N$ and $(2N-3)$ is negligible.) Therefore, we
will only use the $L=1$ and $L=2$ tests, but not higher $L$. Once
the null hypothesis is rejected by either the $L=1$ or $L=2$ test
at a given significance level, then this rejection is correct.
Only the significance level obtained provides a lower limit for
this rejection, because it is still possible that some further
tests with higher $L$ will reject the null hypothesis at a higher
significance level.

There are two problems with the application of this test. One
problem is general and the second is a special problem occurring
for the BATSE data of GRBs.

The first problem is the following. To compare the theoretical
curve with the empirical curve one needs $N$ measured independent
$L$-th nearest distances. We have only one single sample: the
actual distribution of $N$ objects on the sky. In it one has $N$
$L$-th nearest neighbour distances; for any object one calculates
the $L$-th nearest neighbour. But these $N$ distances need not be
independent, because some distances may occur twice. (If the
$L$-th nearest neighbour for $k$-th object is the $m$-th object,
then it may well happen that the $L$-th nearest neighbour for
$m$-th object is the $k$-th object.) Fortunately, this is not an
essential defect excluding its use (\cite{ST89}).

The second problem concerns the case of GRBs alone. It follows
from the fact that the sky is not covered uniformly by the BATSE
instrument. There is a sky-exposure function $g(\delta)$ that
depends on the declination $\delta$ (\cite{mee00}). Therefore, the
theoretical curve from Eq.~(2). and Eq.~(3). is not usable at
once. Fortunately it is easy to take this into account. (In the
previous papers cited in Sect. 1 the effect of a non-uniform
sky-exposure function could also be corrected for. Hence, if we
discuss anisotropy, we always mean intrinsic anisotropy in the
distribution of GRB not caused by the BATSE instrument.)

This correction may be made as follows. Let there be a burst at
declination $\delta$. Then one can always introduce a new
declination $\tilde{\delta}$ unambiguously by the relation
\begin{equation}
2 \int_{-\pi/2}^{\delta} g(\delta') \cos \delta' d\delta' =
A \int_{-\pi/2}^{\tilde{\delta}} \cos \delta' d\delta',
\end{equation}
where
\begin{equation}
A = \int_{-\pi/2}^{\pi/2} g(\delta') \cos \delta' d\delta'.
\end{equation}
This means that - formally - the declination is ``shifted" to a
new value. If there is an intrinsic isotropy in the distribution
of GRBs, the GRBs should also be distributed isotropically in the
new ``shifted" coordinates.  Note that this shift is a standard
method in Statistics (\cite{tw53}, Chapt.1.13). Note also that
this elimination of the non-uniform sky-exposure function is new.

Because one should obtain isotropy in these ``shifted" coordinates
- if there is intrinsic isotropy -  any statistical test that
assumes uniform sky exposure can be used.  Therefore NNA is also
usable. We will provide the NNA test for $L=1$ and $L=2$ in these
``shifted" coordinates for the short, intermediate and long GRBs,
respectively.

\section{Comparison of the method with other tests}

In this section the advantages and disadvantages of NNA are
summarized and compared with other tests mentioned in Sect. 1.

First of all, we want to remark that NNA is highly similar to the
two-point angular correlation function method. In both cases the
key idea is the same: There are $N(N-1)/2$ angular distances among
$N$ objects, and  these measured angular distances are compared
with the theoretically expected distances following from the
random angular distribution of $N$ objects.

Both methods have the great advantage that they are independent of
the choice of coordinate systems and also of other artificial
choices. (For example, in the counts-in-cells method the
boundaries of the cells must be chosen ad hoc (\cite{me00b}). No
such ad hoc choice is needed here.) A further advantage of these
methods lies in the fact that they are also able to detect
eventual anisotropies on small angular scales. (If there are $N$
($N \gg 1$) objects on the sky, then the angular scales - measured
in radians - are large, if these scales are much bigger than
$\simeq \sqrt{4\pi/N}$; small angular scales in radians correspond
with  distances $\simeq \sqrt{4\pi/N}$; for further details see
Scout \& Tout (1989).) This is a great advantage of NNA compared,
e.g., with the counts-in-cells method, which is insensitive to
angular scales much smaller than the cell size (inside a cell
eventual anisotropies may be ``smoothed"; for more details see
M\'esz\'aros et al. (2000b)). In principle, the method based on
spherical harmonics, and  multifractal methods are sensitive to
small scales, too. Nevertheless, on scales $\simeq \sqrt{4\pi/N}$
radians these methods begin to have several technical problems
(see, e.g., M\'esz\'aros et al. (2000a) for further details). All
this means that NNA and the method based on the two-point angular
correlation function may well detect anisotropies on scales
$\simeq \sqrt{4\pi/N}$ radians, which were not detected yet by
other methods.

There are two essential differences between NNA and the method
based on the two-point angular correlation function. The first
concerns the number of distances used: NNA uses only $2N$ angular
distances, but the second method uses $N(N-1)/2$ distances. The
second concerns the procedure of the calculation of significance
levels: The method based on the two-point angular correlation
function needs Monte-Carlo simulations (\cite{me00c}); NNA does
not need them. It is well known that one has to be careful using
pseudo-random generators and Monte-Carlo simulations (see, e.g.,
Chapter 7.0 of Press et al. (1992)). Therefore it is essential to
use a method that does not need these pseudo-random simulations.
On the other hand, this advantage of NNA is lessened by the fact
that it uses only $2N$ distances. This means that NNA may miss
anisotropies detected by the correlation function method. On the
other hand, any anisotropy detected by NNA method {\it should also
be} detected by the correlation function method. In other words,
NNA is not as powerful as the correlation function method. Its
advantage is given mainly by its simplicity and by the fact that
it does not need pseudo-random generations. In any case, its use
in the case of GRBs is certainly justified.

Add to this that, as it is well-known in Statistics (\cite{tw53}),
it is quite usual for different tests to give different
conclusions. (Different tests give different ``trials".) Some
tests may reject the hypothesis of isotropy while further tests do
not reject it. In addition, if two different tests reject it, then
this rejection may occur at different significance levels, too.
Trivially, if there is an isotropic distribution, then no test
should reject the hypothesis of this isotropy. This means that, at
least in principle, one single test rejecting the isotropy is
enough to proclaim the existence of anisotropy. On the other hand,
using several tests is clearly better, because if only one single
test rejects the isotropy one can never exclude with certainty
that there may not have been some unknown technical problems
(pseudo-random simulations, unknown instrumental effects, unknown
systematic errors in measurements, etc...).

In fact this is the situation also for long GRBs (see Sect. 1):
Isotropy of both the short and intermediate subgroups of GRBs is
rejected by several tests, and hence it may be stated that these
subgroups are distributed anisotropically on the sky. On the other
hand, the isotropic distribution of the long-GRB subgroup is
rejected by one single test only; hence its anisotropy is
questionable still.

\section{The results}

\begin{figure}
   \centering
\resizebox{\hsize}{!}
{\includegraphics[angle=-90,width=5cm]{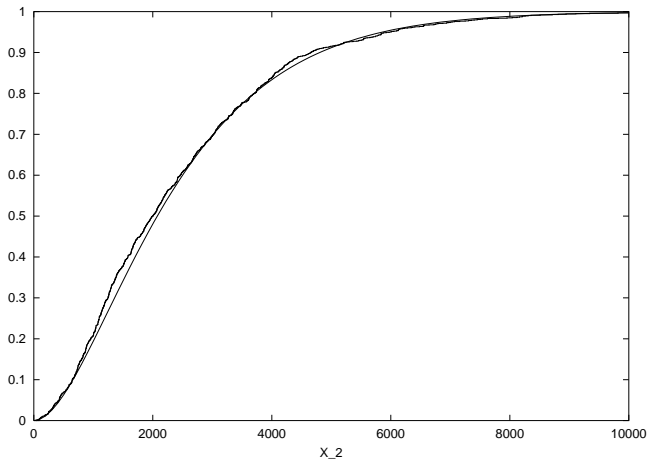}}
      \caption{Kolmogorov-Smirnov test for the second
nearest neighbour for the long subgroup containing $N= 1239$ GRBs.
The theoretical curve is the ``smooth" curve; the empirical curve
is a step function increasing in steps  of $1/1239$. The biggest
difference $D$ between the two curves is at around $X_2 \simeq
1500$, where the theoretical curve is at the bottom.}
   \end{figure}

There are 2702 GRBs in the BATSE Catalog (\cite{mee00}), and from
them
 2037 GRBs have measured $T_{90}$. They are separated
into three subgroups: 497 GRBs having $T_{90} \leq 2$ s comprise
the `` short" subgroup; 301 GRBs having $2$ s $< T_{90} \leq 10$ s
comprise the ``intermediate" subgroup; and 1239 GRBs having
$T_{90}
> 10$ s comprise the ``long" subgroup. Because the existence of the third
intermediate subgroup is not sure yet (see Sect. 1), for safety we
also test the ``non-short" subgroup contaning 1239 + 301 = 1540
GRBs having $T_{90} > 2$ s (i.e. the ``intermediate" and ``long"
subgroups are taken together). For the sake of completeness we
also tested the sample of ``all" GRBs, containing 2702 GRBs. For
any sample  we provide the nearest neighbour and second nearest
neighbour analysis using the standard Kolmogorov-Smirnov test. For
any sample we take the bigger $D$ from the $L=1$ and $L=2$ tests.

Table 1 shows the results.

\begin{table}
\caption{The results of the Kolmogorov-Smirnov tests for the three
subclasses. The first column denotes the subclass; the second the
number of GRBs of this subclass; the third column gives the
difference $D$ between the theoretical and empirical cumulative
function; the fourth column gives the significance (in percentage)
of the rejection of the null-hypothesis of isotropy (Eq. 14.3.9 of
Press et al. (1992)).}
$$
  \begin{array}{rrrr}
 \hline \hline
  subgroup & N & D & \% \\
\hline
  short & 497       & 0.049 &  82.1     \\
  intermediate & 301  &  0.054 & 76.5    \\
  long & 1239  &  0.046 & 99.0    \\
  non-short & 1540 & 0.021 & 51.2 \\
  all      & 2702 & 0.021 & 77.2 \\
   \hline \hline
  \end{array}
$$
\label{tab1}
  \end{table}

Table 1 gives the remarkable result: for the long subclass, and
only for this subclass, is the null-hypothesis of isotropy
rejected at the usual $>95\%$ significance level (viz. 99\%). This
result follows from the $L=2$ test.

Both the theoretical and empirical cumulative distribution
functions for $L=2$ are shown in Fig. 1. From this it follows that
around $X_2 \simeq 1500$, i.e. around $\beta \simeq 0.6$ radians $
\simeq (3-4)$ degrees, the empirical curve lies significantly
above the theoretical one. This means that on this angular scale
the actual angular distribution of GRBs shows an overdensity
compared with the random isotropic case.

\section{Discussion and conclusion}

If one compares the previous tests surveyed in Sect. 1 with Table
1, it seems that the occurrence of anisotropy for the long
subgroup, and for this subgroup only, is a surprising result.

Nevertheless, nothing unexpected is occurring here. First, NNA is
sensitive on small angular scales (for the long subgroup on
$\simeq \sqrt{4\pi/N} = 0.2$ radians). Previous tests were
sensitive on large angular scales, so  it is not strange to detect
anisotropy on angular scales of a few degrees.  Second, the
two-point angular correlation function shows  anisotropy for this
subgroup, too (see the survey in Sect. 1). (As is noted in Sect.
3, it is to be expected that  anisotropy detected by NNA should be
detected by the method based on the correlation function, too; the
opposite does not hold). Third, from Eq. 14.3.9 of Press et al.
(1992) it follows that the significance level  increases with
$D\sqrt{N}$. From Table 1 one sees that $D$ is more or less the
same for the first three subclasses; hence, the long subclass is
anisotropic due to high $N$. It is therefore quite possible that
the anisotropies for short and intermediate subclasses
respectively are not detected by NNA due to the small $N$. Fourth,
visual inspection of the distribution of long GRBs on the sky
(Fig. 2) also shows some grouping on small angular scales. Fifth,
in fact some anisotropy on degree scales is expected from
Cosmology: at $z \ll 1$ ($z$ is the redshift)  the distribution of
galaxies and other objects is inhomogeneous on scales $\simeq
(10-100)$ Mpc (see \cite{me97} and references therein). At high
$z$, where the long GRBs dominantly are, these scales correspond
to angular scales of some degrees. (Direct measurements from GRB
afterglows give $z = 4.5$ for the maximal redshift (\cite{me01});
indirect observational data also allow $z \simeq 20$
(\cite{mm96}).) Hence, if the present-day inhomogeneities exist
also at the high redshifts and if the distribution of long GRBs
reflects the distribution of matter at these high redshifts, then
anisotropy of long GRBs on degree scales is quite possible.

All this means that the occurrence of anisotropy for the long
subgroup and the simultaneous non-occurrence of anisotropy for the
intermediate and short subgroup is not strange. On the other hand,
the isotropy of the non-short sample is remarkable. It seems that
GRBs from intermediate and long subgroups separately are
anisotropically distributed; but differently and that their
anisotropies ``cancel". It is urgent to clarify the status of the
intermediate subgroup - namely whether is it a real different
subgroup or not (see Sect. 1). Our result, together with
Horv\'ath's recent results (Horv\'ath 2002), suggest that the
intermediate subclass may be a real subgroup.

The result concerning the sample denoted by ``all" has no real
meaning, because the short and long subclasses may be different
phenomena (see Sect. 1).

   \begin{figure*}
   \centering
\resizebox{\hsize}{!}
{\includegraphics[angle=-90,width=3cm]{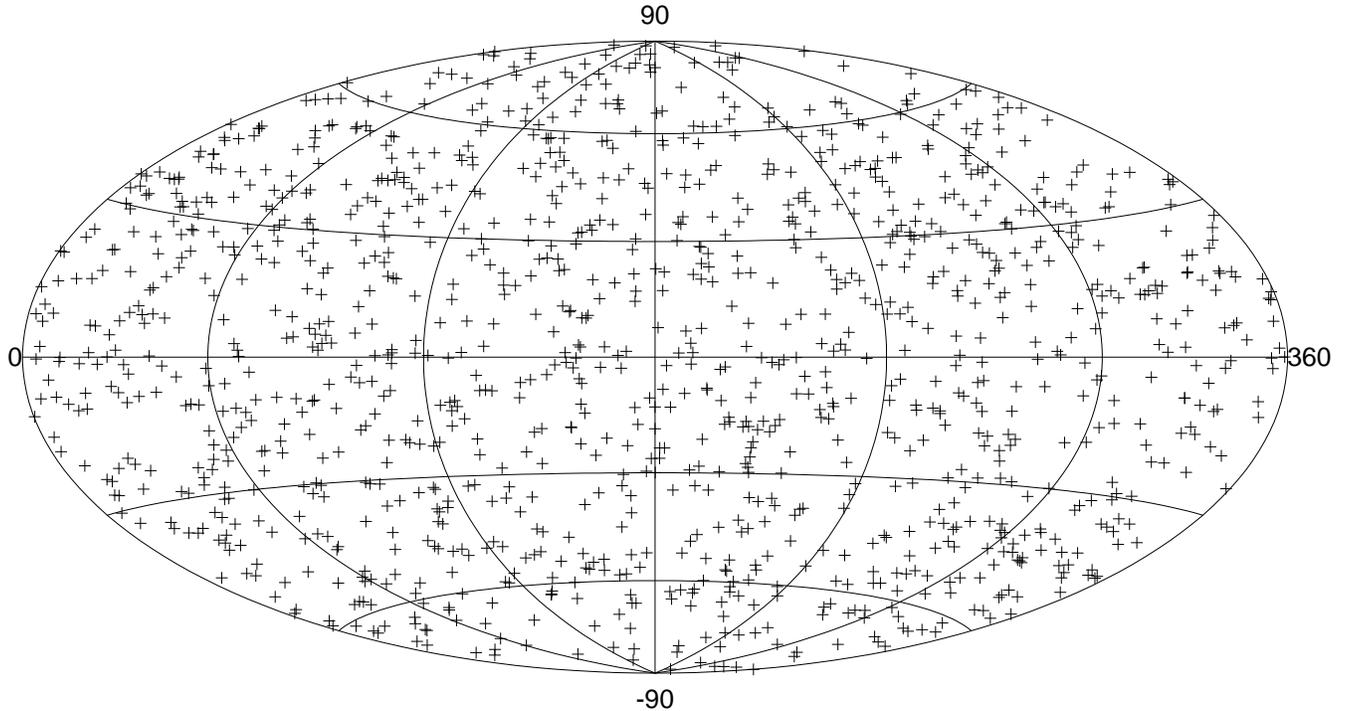}}
      \caption{Distribution of 1239 long GRBs on the sky in equatorial
      coordinates and Aitof projection.}
   \end{figure*}

As was already noted, theoretically (see Sect. 3), it is
sufficient to have one single rejection of the null hypothesis
from one single statistical test. In practice, the result is more
reliable if several tests reject the null hypthesis. Concerning
the long subgroup this is what we get here, because both NNA and
the correlation function method reject the isotropy. Hence this paper
may lead to the conclusion that the long bursts are also
distributed anisotropically.

Nevertheless, the positive result of this paper, together with a
similar result from the angular correlation function is
encouraging, but not yet definite; we still have to be very
cautious, for a number of reasons:
\begin{itemize}
\item These two methods use the same idea: the
measured angular distances are compared with what would be
theoretically expected for an isotropic sample. Hence, these two
methods cannot be considered as fully independent statistical
tests.
\item As was noted, the use of NNA has a general problem, 
because some distances
may occur twice. Hence, in principle, the results of this test
alone need not be final.
\item The rejection of isotropy on  small angular 
scales may by caused - at least in
principle - by a wholly different phenomenon; even if the long
bursts are distributed isotropically, but some bursts are
occurring at the same position several times (i.e. if some bursts
are repeating), then the NNA test may give a positive result. This
mixing of two effects is discussed, e.g., by Brainerd (1996).
Nowadays it is practically certain that for the long bursts no
repetition occurs. This follows mainly from the models of these
GRBs - they always assume a total destruction of source (see, e.g,
M\'esz\'aros P. (2001) for a survey of models). On the other hand,
artificial instrumental effects cannot be excluded yet for the
long bursts. It is in principle possible that a long burst is
detected {\it again} by the BATSE instrument after one or two
orbits of the satellite. Then the new detection would be included
in the Catalog as a new burst, but one would have in fact one
single burst (V. Connaughton, private communication). Hence,
strictly from a statistical point of view, instrumental repetition
is not excluded yet definitely.
   In order to discuss this possibility we have searched for  pairs of
GRBs in the sample denoted ``all", in which: A. the later GRB
occurred  4 hours after the first event, B. the angular distance
between them is less  than 10 degree, C. at least one GRB from the
pair should have $T_{90} > 10$ s. Only three pairs were found that
fulfilled all these requirements. Their BATSE trigger numbers are:
5648-5649, 6165-6166 and 7359-7360. In addition, from these six
GRBs only three  (5648, 6165, 7360) belong to the long subclass;
hence, there is no pair in which both GRBs belong to the long
subgroup. It is highly questionable that in these three cases
instrumental repetition occurs. But, even assuming this, if one
deletes three GRBs from the ``long" sample containing 1239 GRBs,
it does not make an essential change in the sample: the effect of
instrumental repetition can hardly  have any importance for our
conclusions.
\item The non-short sample does not show
anisotropy. This means that, if the existence of a third
intermediate subgroup were not confirmed in the future (this is
still possible even after Horv\'ath's recent paper (\cite{ho02})),
then the anisotropy obtained in this paper would be an interesting
and remarkable result but not a proof of the anisotropy of a
separate physically well-defined subclass of GRBs.
\item The positional errors of GRBs are 
large (a few degrees), and are comparable
with the angular scale of the expected anisotropy. Nevertheless,
an earlier study (\cite{te96}) shows that these positional errors
 cancel  on average, and hence their effect need not
be important.
\end{itemize}

Keeping all this in mind, we conclude that {\it the anisotropy of
the long subclass at a $\geq 99\%$ significance level may be
real.} However, to reach a more definite conclusion several
further independent tests are still needed.

We are aiming to provide such tests on BATSE data in future. We
also hope that the results of this paper - together with those of
the earlier papers - will also encourage others to provide
independent statistical tests on the angular distribution of GRBs.

\begin{acknowledgements}
Useful discussions with Drs. Z. Bagoly, L.G. Bal\'azs, I. Horv\'ath,
P. M\'esz\'aros, R. Vavrek and the valuable remarks of the
referee, Dr. V. Connaughton, are kindly acknowledged.
This research was supported by Czech Research Grant J13/98: 113200004.
\end{acknowledgements}

\end{document}